\documentclass[rnote]{aa} 
\usepackage{graphicx}
\usepackage{natbib}
\usepackage{txfonts}
\usepackage{deluxetable}
\begin{document}
   \title{Adaptive optics observations of the T10 ultracool dwarf UGPS~J072227.51-054031.2\thanks{Based on observations obtained with the ESO Telescopes at the Paranal Observatory under programme 285.C-5004}}

   \author{H. Bouy\inst{1}
          \and J. Girard\inst{2}
          \and E.~L. Mart\'\i n\inst{1}
          \and N. Hu\'elamo\inst{1}
          \and P.~W. Lucas\inst{3}
          }

   \institute{Centro de Astrobiolog\'\i a, INTA-CSIC, PO BOX 78, E-28691, Villanueva de la Ca\~nada, Madrid, Spain\\
     \and
     European Southern Observatory, Alonso de Cordova 3107, Vitacura, Casilla 19001, Santiago 19, Chile \\
     \and
     Centre for Astrophysics, STRI, University of Hertfordshire, College Lane, Hatfield AL10 9AB, UK.\\
             }

   \date{Received ...; accepted ...}

 
  \abstract
   {With a spectral type of T10, UGPS~J072227.51-054031.2 is one of the coolest objects known to date in the solar neighborhood.}
   {Multiple systems are relatively common among early and mid-T dwarfs. We search for faint and close companions around UGPS~J072227.51-054031.2.  }
   {We have obtained high spatial resolution images in the $H$ and $K_{s}$ bands using adaptive optics at the Very Large Telescope.}
   {With a Strehl ratio in the range 10--15\%, the final images allow us to rule out the presence of a companion brighter than $H\lesssim$19.4~mag at separation larger than 50~mas, and $H\lesssim$21.4~mag at separation larger than 0\farcs1.}
   {}


   \maketitle
%

\section{Introduction}
Tremendous efforts have been made over the past decade to study the ultracool end of the local field mass function. To date, several hundreds methane dwarfs have been discovered as a result of sensitive surveys such as the 2MASS \citep{2006AJ....131.1163S}, DENIS \citep{1997Msngr..87...27E}, UKIDSS \citep{UKIDSS} and CFHTBDS \citep{2008A&A...484..469D} near-infrared surveys, and the SDSS survey \citep{2000AJ...120..1579}. As methane dwarfs cannot sustain fusion in their core, their temperature drops continuously with time. A number of studies have therefore searched for ever cooler objects, representative of the proposed Y spectral class, \citep{2008A&A...484..469D,2008MNRAS.391..320B,2007MNRAS.381.1400W} and encountered a few objects with effective temperatures of only $\approx$600~K. Most recently, \citet{2010arXiv1004.0317L} reported the discovery of an ultracool dwarf (UGPS~J072227.51-054031.2) with an effective temperature of only T$_{\rm eff}\approx$500~K, making it the coolest brown dwarf known to date. They classify it as a T10, at the boundary between T and Y dwarfs.  Since methane dwarfs are frequently found in multiple systems \citep{2003ApJ...586..512B, 2008A&A...490..763G}, we decided to perform high angular resolution observations of the target to look for close companions. Resolving the source into a multiple system would extend further the domain of absolute magnitude for known ultracool dwarfs, and might significantly revise the estimated luminosity and mass of the primary component.


\section{Observations and data reduction}
UGPS~J072227.51-054031.2 was observed with NACO, the Very Large Telescope (VLT) adaptive optics system \citep{2003SPIE.4841..944L,2003SPIE.4839..140R} and its laser guide star on the 12$^{\rm th}$ of April 2010 as part of program 285.C-5004. A set of deep images were obtained in $H$ and $K_{s}$ where the compromise between the target's luminosity and the AO performances is the best.

\begin{figure}
    \centering
   \includegraphics[width=0.45\textwidth]{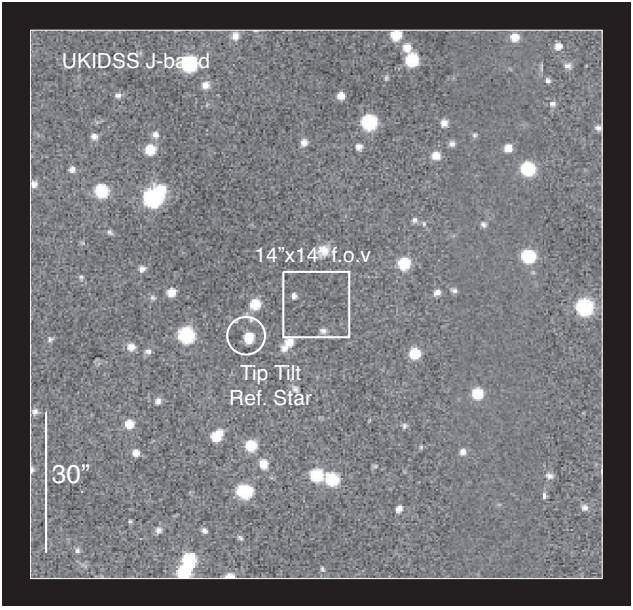}
      \caption{UKIDSS $J$-band image showing the field of view of the NACO images, and the tip-tilt reference star. The scale is indicated. North is up and east is left. }
         \label{ukidss}
   \end{figure}

At the time the observations were prepared, \citet{2010arXiv1004.0317L} had ruled out the presence of companion at distances larger than 3\arcsec\, using deep seeing limited images. We therefore decided to limit our study to the immediate vicinity of the target, and used a 512$\times$512 pixels window with the S27 camera, providing a final field of view of $\approx$14$\times$14\arcsec. This setting allows a faster read-out and an optimized use of the cube mode. In cube mode, a data-cube with each individual \emph{DIT} frame is saved, allowing a careful frame selection and optimized weighting in the post-processing and co-addition. A set of 208 and 240 4~s individual images were acquired in $H$ and $K_{s}$, respectively. The target was too faint to be used as reference star for NACO's visible or near-infrared wavefront sensors. We therefore used the laser guide star for the high order corrections, and the $R\sim$15.8~mag star 2MASS~J07222849-0540377 located at $\sim$19\arcsec\, from the science target for tip-tilt correction (see Fig~\ref{ukidss}). The ESO seeing monitor reports a DIMM seeing between 0\farcs5$\lesssim \sigma \lesssim$1\farcs0, a coherence time of the atmospheric turbulence between 5$\lesssim \tau_{0} \lesssim$10~ms, while the sky transparency was classified as photometric. 

\begin{figure}
    \centering
   \includegraphics[width=0.45\textwidth]{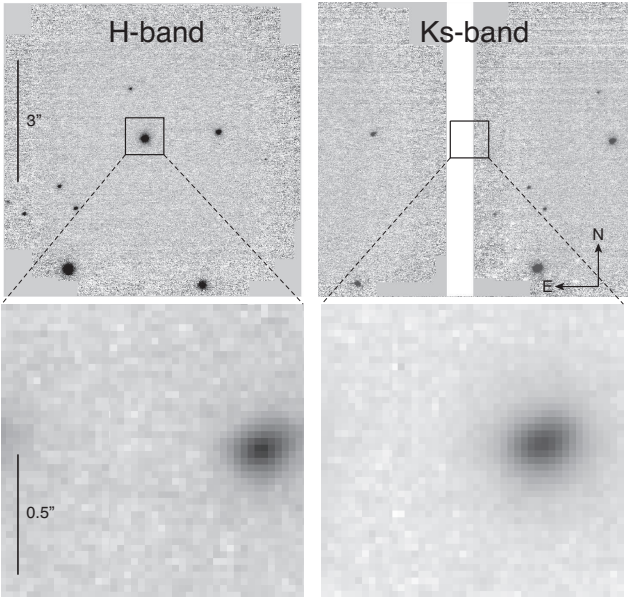}
      \caption{NACO $H$ (left) and $K_{s}$ (right) images of UGP~ J072227.51-054031.2. The lower panels show a zoom on the target. The scale and orientation are indicated. }
         \label{images}
   \end{figure}

The individual images were successively dark subtracted and flatfielded using calibration frames obtained as part of the standard calibration sequence. The Strehl ratio was then measured on the target for each individual image of the cube by fitting a Moffat function to the target's point spread function (PSF), and dividing the corresponding normalized peak to the normalized peak of a theoretical VLT diffraction limited PSF. The individual images were then co-added weighted by their respective Strehl ratio. The final Strehl ratio is $\approx$12\% in both the $H$ and $K_{s}$ image. Figure~\ref{images} shows the final images.

We extracted the photometry of all sources brighter than the 3-$\sigma$ local noise using standard aperture photometry procedures within IRAF\footnote{IRAF is distributed by National Optical Astronomy Observatories, which is operated by the Association of Universities for Research in Astronomy, Inc., under contract with the National Science Foundation.}. The zeropoints were computed using the H and K$_{s}$ photometry of 4 sources with a counterpart in the UKIDSS survey. The final uncertainties are largely dominated by the zeropoint uncertainties, computed as the 1-$\sigma$ standard deviation between the 4 measurements.

\section{Analysis and conclusions}
\begin{figure}
   \centering
   \includegraphics[width=0.45\textwidth]{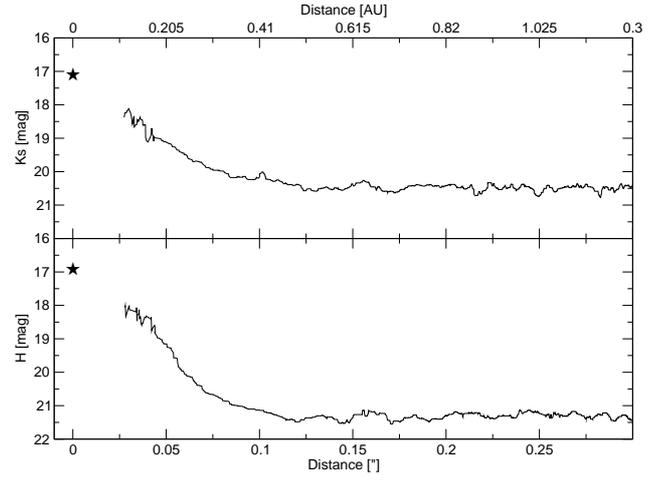}
      \caption{Limit of sensitivity of our observations, computed as the 3-$\sigma$ standard deviation of the radial profile of the PSF. The projected distance scale assuming a distance of 4.1~pc is indicated. The luminosity of UGPS~J072227.51-054031.2 is indicated with a star.}
         \label{sensitivity}
   \end{figure}

No obvious close companion is detected around the target. A careful PSF subtraction using the 2 brightest sources present in the field produces clean residuals. Figure~\ref{sensitivity} shows the 3-$\sigma$ limit of detection of the $H$ and $K_{s}$ images, computed using the standard deviation of the radial profile of the PSF. The source is slightly elongated towards the tip-tilt reference star, and Figure~\ref{sensitivity} represents an average of the limit of sensitivity over all azimuthal directions. The observations allow us to rule out the presence of a companion brighter than $H\lesssim$19.2~mag and $K\lesssim$19.1~mag at separation larger than 50~mas (0.205~AU at 4.1~pc), and $H\lesssim$21.2~mag and $K_{s}\lesssim$20.2~mag at separation larger than 0\farcs1 (0.41~AU at 4.1~pc). A total of 7 additional sources are present in the image. Table~\ref{phot} gives an overview of their astrometry and photometry. All but one are detected in the J-band image dated 28 November 2006 of \citet{2010arXiv1004.0317L}, and are located at the same absolute position (within 0\farcs1) in our NACO images while the target has moved by more than 3\arcsec\, between these two epochs. We therefore rule out these 6 sources as possible companions based on their inconsistent proper motion. The last source is significantly redder than the target (H-K$_{s}$=0.36~mag). As a cooler companion would be expected to be bluer, we reject this last source as a possible companion. 

\begin{table}
\caption[]{Astrometry and photometry of the sources present in the NACO images} 
\label{phot}
\begin{tabular}{lccccccc}
\hline
RA (J2000) & Dec (J2000) & Z & Y & J & H & K$_{s}$ \\
\hline
  07:22:27.60 & -05:40:38.3 & 17.24$\pm$ 0.01 & 16.96$\pm$0.01 & 16.46$\pm$0.01 & 15.77$\pm$0.02 & 15.71$\pm$0.15 \\
  07:22:27.28 & -05:40:30.0 & 20.45$\pm$ 0.06 & 17.37$\pm$0.01 & 16.49$\pm$0.01 & 16.87$\pm$0.02 & 16.72$\pm$0.15 \\
  07:22:27.02 & -05:40:39.2 & 17.83$\pm$ 0.01 & 17.69$\pm$0.01 & 17.32$\pm$0.01 & 17.00$\pm$0.02 & 17.25$\pm$0.15 \\
  07:22:26.97 & -05:40:29.4 & 19.66$\pm$ 0.03 & 19.04$\pm$0.04 & 18.52$\pm$0.03 & 18.13$\pm$0.02 & 17.85$\pm$0.15 \\
  07:22:27.79 & -05:40:34.7 & 20.86$\pm$ 0.08 & 21.51$\pm$0.35 & 19.70$\pm$0.09 & 19.24$\pm$0.02 & \nodata    \\
  07:22:27.64 & -05:40:33.0 & 21.20$\pm$ 0.11 & 20.34$\pm$0.13 & 19.74$\pm$0.10 & 19.36$\pm$0.02 & 18.86$\pm$0.15 \\
  07:22:27.56 & -05:40:34.4 & 20.80$\pm$ 0.08 & 20.65$\pm$0.17 & 19.82$\pm$0.10 & 19.18$\pm$0.02 & 18.99$\pm$0.15 \\
  07:22:27.34 & -05:40:26.8 & 21.48$\pm$ 0.14 & \nodata        & \nodata        & 19.90$\pm$0.03 & 19.54$\pm$0.15\\
\hline
\end{tabular}
Note: Z,Y and J photometry from \citep{2010arXiv1004.0317L}; H and K$_{s}$ photometry from NACO (this study).
\end{table}

 
\bibliographystyle{aa}

\end{document}